\documentstyle[12pt]{article}

\font\twlgot =eufm10 scaled \magstep1 \font\egtgot =eufm8
\font\sevgot =eufm7 \font\twlmsb =msbm10 scaled \magstep1
\font\egtmsb =msbm8 \font\sevmsb =msbm7

\newfam\gotfam

\textfont\gotfam\twlgot \scriptfont\gotfam\egtgot
\scriptscriptfont\gotfam\sevgot

\newfam\msbfam
\textfont\msbfam\twlmsb \scriptfont\msbfam\egtmsb
\scriptscriptfont\msbfam\sevmsb
\def\Bbb{\protect\pBbb}
\def\pBbb{\relax\ifmmode\expandafter\Bb\else\typeout{You cann't use
Bbb in text mode}\fi}
\def\Bb #1{{\fam\msbfam\relax#1}}

\def\thebibliography#1{\section*{References}\list
  {[\arabic{enumi}]}{\settowidth\labelwidth{#1}\leftmargin\labelwidth
    \advance\leftmargin\labelsep
    \usecounter{enumi}}
    \def\newblock{\hskip .11em plus .33em minus .07em}
    \sloppy\clubpenalty4000\widowpenalty4000
    \sfcode`\.=1000\relax}

\def\op#1{\mathop{\fam0 #1}\limits}

\newcommand{\beq}{\begin{equation}}
\newcommand{\eeq}{\end{equation}}
\newcommand{\ben}{\begin{eqnarray}}
\newcommand{\een}{\end{eqnarray}}
\newcommand{\be}{\begin{eqnarray*}}
\newcommand{\ee}{\end{eqnarray*}}
\newcommand{\bea}{\begin{eqalph}}
\newcommand{\eea}{\end{eqalph}}

\newcommand{\cD}{{\cal D}}

\newcommand{\cL}{{\cal L}}

\newcommand{\cK}{{\cal K}}
\newcommand{\cU}{{\cal U}}

\newcommand{\bb}{{\bf 1}}

\newcommand{\al}{\alpha}

\newcommand{\dl}{\delta}
\newcommand{\la}{\lambda}

\newcommand{\f}{\phi}

\newcommand{\m}{\mu}

\newcommand{\vf}{\varphi}

\newcommand{\lng}{\langle}
\newcommand{\rng}{\rangle}

\newcommand{\wt}{\widetilde}
\newcommand{\wh}{\widehat}
\newcommand{\ol}{\overline}
\newcommand{\dr}{\partial}
\newcommand{\ar}{\op\longrightarrow}

\newcommand{\ve}{\varepsilon}

\newcounter{eqalph}
\newcounter{equationa}
\newcounter{remark}
\newcounter{example}
\newcounter{theorem}
\newcounter{proposition}
\newcounter{lemma}
\newcounter{corollary}
\newcounter{definition}
\def\theremark{\arabic{remark}}
\def\thetheorem{\arabic{theorem}}

\def\thedefinition{\arabic{definition}}

\newenvironment{prop}{\refstepcounter{theorem}
\bigskip\noindent{\bf Proposition \thetheorem.}\it}{\medskip}
\newenvironment{lem}{\refstepcounter{theorem}
\bigskip\noindent{\bf Lemma \thetheorem.}\it}{\medskip}

\newenvironment{eqalph}{\stepcounter{equation}
\setcounter{equationa}{\value{equation}} \setcounter{equation}{0}

\begin{eqnarray}}{\end{eqnarray}
\setcounter{equation}{\value{equationa}}}

\newcommand{\mar}[1]{}

\begin{document}
\hbox{}

{\parindent=0pt

{\large \bf On the mathematical origin of quantum space-time}
\bigskip

{\sc G.Sardanashvily}\footnote{E-mail:
gennadi.sardanashvily@unicam.it}

{\sl Department of Theoretical Physics, Moscow State University,
117234 Moscow, Russia}

\bigskip
\bigskip

\begin{small}

{\bf Abstract} An Euclidean topological space $E$ is homeomorphic
to the subset of $\dl$-functions of the space $\cD'(E)$ of
Schwartz distributions on $E$. Herewith, any smooth function of
compact support on $E$ is extended onto $\cD'(E)$. One can think
of these extensions as sui generis quantum deformations. In
quantum models, one therefore should replace integration of
functions over $E$ with that over $\cD'(E)$.

\end{small}

 }

\bigskip
\bigskip

A space-time in field theory, except noncommutative field theory,
is traditionally described as a finite-dimensional smooth
manifold, locally homeomorphic to an Euclidean topological space
$E=\Bbb R^n$. The following fact (Proposition 1) enables us to
think that a space-time might be a wider space of Schwartz
distributions on $E$.

Let $E=\Bbb R^n$ be an Euclidean topological space. Let $\cD(E)$
be the nuclear space of smooth complex functions of compact
support on $E$. Its topological dual $\cD'(E)$ is the space of
Schwartz distributions on $E$, provided with the weak$^*$ topology
\cite{trev,bogol}. Since $\cD(E)$ is reflexive and the strong
topology on $\cD'(E)$ is equivalent to the weak$^*$ one, $\cD(E)$
is the a topological dual of $\cD'(E)$. Therefore, any continuous
form on $\cD'(E)$ is completely determined by its restriction
\be
\lng\f,\dl_x\rng=\int \f(x')\dl(x-x')d^nx=\f(x), \qquad x\in E,
\ee
to the subset $T_\dl(E)\subset \cD'(E)$ of $\dl$-functions.

\begin{prop} \label{f3} \mar{f3}
The assignment
\mar{f2}\beq
s_\dl:E\ni x\to \dl_x\in \cD'(E) \label{f2}
\eeq
is a homeomorphism of $E$ onto the subset $T_\dl(E)\subset
\cD'(E)$ of $\dl$-functions endowed with the relative topology
(see Appendix for the proof).
\end{prop}

As a consequence,  $T_\dl(E)$ is isomorphic to the topological
vector space $E$ with respect to the operations $\dl_x \oplus
\dl_{x'}= \dl_{x+x'}$, $\la\odot \dl_x=\dl_{\la x}$. Moreover, the
injection $E\to T_\dl(E)\subset D'(E)$ is smooth \cite{col}.
Therefore, we can identify $E$ with a topological subspace
$E=T_\dl(E)$ of the space of Schwartz distributions. Herewith, any
smooth function $\f$ of compact support on $E= T_\dl(E)$ is
extended to a continuous form
\mar{f30}\beq
\wt \f(w)=\lng\f,w\rng, \qquad w\in \cD'(E), \label{f30}
\eeq
on the space of Schwartz distributions $\cD'(E)$. One can think of
this extension as being a quantum deformation of $\f$ as follows.

The space $\cD(E)$ is a dense subset of the Schwartz space $S(E)$
of smooth complex functions of rapid decrease on $E$. Moreover,
the injection $\cD(E)\to S(E)$ is continuous. The topological dual
of $S(E)$ is the space $S'(E)$ of tempered distributions, which is
a subset of the space $\cD'(E)$ of Schwartz distributions. In QFT,
one considers the Borchers algebra
\mar{x2}\beq
A_S=\Bbb C\oplus S(E) \oplus S(E\oplus E)\oplus \cdots \oplus
S(\op\oplus^kE) \oplus\cdots, \label{x2}
\eeq
treated as a quantum algebra of scalar fields \cite{borch,hor}.
Being provided with the inductive limit topology, the algebra
$A_S$ (\ref{x2}) is an involutive nuclear barreled LF-algebra
\cite{belang}. It follows that a linear form $f$ on $A_S$ is
continuous iff its restriction $f_k$ to each $S(\op\oplus^k E)$ is
well \cite{trev}. Therefore any continuous positive form on $A_S$
is represented by a family of tempered distributions $W_k\in
S'(\op\oplus^kE)$, $k=1,\ldots,$ such that
\mar{qm810}\beq
f_k(\f(x_1,\ldots,x_k))= \int
W_k(x_1,\ldots,x_k)\f(x_1,\ldots,x_k) d^nx_1\cdots d^nx_k, \qquad
\f\in S(\op\oplus^kE). \label{qm810}
\eeq
For instance, the states of scalar quantum fields on the Minkowski
space $\Bbb R^4$ are described by the Wightman functions
$W_k\subset S'(\Bbb R^{4k})$ \cite{bogol}.

Any state of $A_S$ is also a state of its subalgebra
\be
A_\cD=\Bbb C\oplus \cD(E) \oplus \cD(E\oplus E)\oplus \cdots
\oplus \cD(\op\oplus^kE) \oplus\cdots.
\ee
This quantization can be treated as follows. Given a function
$\f\in \cD(\op\oplus^k E)$ on $\op\oplus^k E$, we have its quantum
deformation
\mar{f19}\beq
\wh \f= \f +f_k(\f)\subset C^\infty(\op\oplus^k E).\label{f19}
\eeq
Let $\op\oplus^k E$ be identified to the subspace
$T_\dl(\op\oplus^k E)\subset \cD'(\op\oplus^k E)$ of
$\dl$-functions on $\op\oplus^k E$. Then the quantum deformation
$\wh\f$ (\ref{f19}) of $\f$ comes from the extension of $\f$ onto
$\cD'(\op\oplus^k E)$ by the formula
\be
\wh \f(z)=\f(z+W_k), \qquad  z+W_k\in S'(\op\oplus^k E)\subset
\cD'(\op\oplus^k E).
\ee

Generalizing this construction, let us consider a continuous
injection
\be
s: \op\oplus^k E\ni z\to s_z\in \cD'(\op\oplus^kE)
\ee
and a continuous function
\be
s_\f:\op\oplus^kE\ni z\to s_z(\f) \in\Bbb C.
\ee
for any $\f\in \cD(\op\oplus^kE)$. For instance, the map $s_\dl$
(\ref{f2}) where $s_{\dl,\f}=\f$ is of this type. Given a function
$\f\in \cD(\op\oplus^kE)$, we agree to call
\mar{f20}\beq
\wh \f= \f +s_\f, \qquad \wh \f(z)=\f(z) + s_z(\f)= \f(z+s_z)
\label{f20}
\eeq
the quantum deformation of $\f$ and to treat it as a function on
the quantum space $\wh E=(s_\dl + s)(E)\subset \cD'(E)$.

For instance, let $\f(x,y)\in \cD(E\oplus E)$ be a symmetric
function on $E\oplus E$. Then its quantum deformation (\ref{f20})
obeys the commutation relation
\be
\wh \f(x,y)-\wh \f(y,x)=\lng\f, s_{x,y}-s_{y,x}\rng.
\ee

Let $E$ be coordinated by $(x^\la)$, and let us consider a
function $x^1x^2$ on $E$, though it is not of compact support. Let
us choose a map $s$ such that all distributions $s_x$, $x\in E$,
are of compact support. Its quantum deformation is
$\wh{x^1x^2}=x^1x^2 + s_x(x^1x^2)$. It is readily observed that
$\wh{x^1x^2}-\wh{x^2x^1}=0$, i.e., coordinates on a quantum space
commute with each other, in contrast to a space in noncommutative
field theory.

Bearing in mind quantum deformations $\wh\f$ (\ref{f30}) of
functions $\f$ on $E$, one should replace integration of functions
over $E$ with that over $\cD'(E)$. Here, we summarize the relevant
material on integration over the space of Schwartz distributions
$\cD'(E)$.
\bigskip

{\bf I.} Due to the homeomorphism (\ref{f2}), the space $T_\dl(E)$
is provided with the measure $d^nx$, invariant with respect to
translations $\dl_x\to \dl_{x+a}$.

\bigskip

{\bf II.} The space $M(E, \Bbb C)$ of measures on $E$ is the
topological dual of the space $\cK(E,\Bbb C)$ of continuous
functions of compact support on $E$ endowed with the inductive
limit topology (see Appendix). The space $M(E,\Bbb C)$ is provided
with the weak$^*$ topology. It is homeomorphic to a subspace of
$\cD'(E)$ provided with the relative topology. It follows that,
for any measure $\nu$ on $E$, there exists an element $w_\nu\in
\cD'(E)$ and the Dirac measure $\ve_\nu$ of support at $w_\nu$
such that, for each $\f\in\cD(E)$, we have
\be
\op\int_E \f\nu(x) = \lng\f,w_\nu\rng=\op\int_{\cD'(E)}
\lng\f,w\rng \ve_\nu(w).
\ee
Let $T_x \subset \cD'(E)$ denote a subspace of point measures
$\la\dl_x$, $\la\in \Bbb C$, on $E=T_\dl(E)$. It is a Banach space
with respect to the norm $||\la\dl_x||=|\la|$. Let us consider the
direct product
\mar{f16}\beq
T(E)=\op\prod_{x\in E} T_x. \label{f16}
\eeq
By analogy with the notion of a Hilbert integral \cite{dixm}, we
define the Banach space integral $(T(E), \cL(E), d^nx)$ where
$\cL(E)$ is a set of fields
\be
\wh \vf: E\ni x\to \vf_x\dl_x\in T(E)
\ee
such that:

$\bullet$ the range of $\cL(E)$ is a vector subspace of the direct
product $T(E)$ (\ref{f16});

$\bullet$ there is a countable set $\{\vf^i\}$ of elements of
$\cL(E)$ such that, for any $x\in E$, the set $\{\vf^i_x\}$ is
total in $T_x$;

$\bullet$ the function $x\to ||\vf_x||=|\vf_x|$ is
$d^nx$-integrable for any $\vf\in\cL(E)$.

\noindent Let $\cL(E)=L^2(E, d^nx)$ be the space of complex square
$d^nx$-integrable functions on $E$. Clearly, $\cD(E)\subset
\cL(E)$, and there is an injection $\cL(E)\to M(E,\Bbb C)\subset
\cD'(E)$ such that
\be
\vf(\f)=\int \f(x')\vf_x\dl(x'-x)d^nx d^nx'.
\ee
Therefore, let
\be
\int \f_x\dl_x d^nx
\ee
denote the image of $\vf$ in $\cD'(E)$. Then any $d^nx$-equivalent
measure $\nu=c^2d^nx$ (where $c\in L^2(E,d^nx)$ is strictly
positive almost everywhere on $E$) defines the corresponding
element
\be
w_\nu=\int c^2(x)\dl_x d^nx
\ee
of $\cD'(E)$. For instance, if $\nu=d^nx$, we have $\vf_x=1$ and
\be
w_\nu=\int \dl_x d^nx.
\ee

\bigskip

{\bf III.} Let $Q$ be an arbitrary nuclear space (e.g., $\cD(E)$,
$S(E)$) and $Q'$ its topological dual (e.g., $\cD'(E)$, $S'(E)$).
A complex function $Z(q)$ on $Q$ is called positive-definite if
$Z(0)=1$ and
\be
\op\sum_{i,j} Z(q_i-q_j)\ol \la_i \la_j\geq 0
\ee
for any finite set $q_1,\ldots,q_m$ of elements of $Q$ and
arbitrary complex numbers $\la_1,\ldots,\la_m$. In accordance with
the well-known  Bochner theorem for nuclear spaces
\cite{bochn,gelf64,bourb6}, any continuous positive-definite
function $Z(q)$ on a nuclear space $Q$ is the Fourier transform
\mar{qm545}\beq
Z(q)=\int\exp[i\lng q,w\rng]\m(w) \label{qm545}
\eeq
of a positive measure $\m$ of total mass 1 on the dual $Q'$ of
$Q$, and {\it vice versa}.

Note that there is no translationally-invariant measure on $Q'$.
Let a nuclear space $Q$ be provided with a separately continuous
non-degenerate Hermitian form $\lng.|.\rng$. In the case of
$Q=\cD(E)$, we have
\be
\lng\f|\f'\rng=\int \f \ol \f'd^nx.
\ee
Let $w_q$, $q\in Q$, be an element of $Q'$ given by the condition
$\lng q',w_q\rng=\lng q'|q\rng$ for all $q'\in Q$. These elements
form the image of the monomorphism $Q\to Q'$ determined by the
Hermitian form $\lng.|.\rng$ on $Q$. If a measure $\m$ in
(\ref{qm545}) remains equivalent under translations
\be
Q'\ni w\mapsto w+w_q \in Q', \qquad  \forall w_q\in Q\subset Q',
\ee
in $Q'$, it is called translationally quasi-invariant. However, it
does not remains equivalent under an arbitrary translation in
$Q'$, unless $Q$ is finite-dimensional.

Gaussian measures exemplify translationally quasi-invariant
measures on the dual $Q'$ of a nuclear space $Q$. The Fourier
transform of a Gaussian measure reads
\be
Z(q)=\exp\left[-\frac12 B(q)\right],
\ee
where $B(q)$ is a seminorm on $Q'$ called the covariance form. Let
$\m_K$ be a Gaussian measure on $Q'$ whose Fourier transform
\be
Z_K(q)=\exp[-\frac12 B_K(q)]
\ee
is characterized by the covariance form $B_K(q)=\lng
K^{-1}q|K^{-1}q\rng$, where $K$ is a bounded invertible operator
in the Hilbert completion $\wt Q$ of $Q$ with respect to the
Hermitian form $\lng.|.\rng$. The Gaussian measure $\m_K$ is
translationally quasi-invariant. It is equivalent $\m$ if
\be
{\rm Tr}(\bb -\frac12 KK^+)<\infty.
\ee
For instance, the Gaussian measures $\m$ and $\m'$ possessing the
Fourier transforms
\be
Z(q)=\exp[-\la^2\lng q|q\rng], \qquad Z(q)=\exp[-\la'^2\lng
q|q\rng] \qquad \la,\la'\in\Bbb R,
\ee
are not equivalent if $\la\neq \la'$.

If the function $\Bbb R\ni t\to Z(tq)$ is analytic on $\Bbb R$ at
$t=0$ for all $q\in Q$, then one can show that the function $\lng
q|u\rng$ on $Q'$ (e.g., the extension $\wt\f$ (\ref{f30}) of $\f$
onto $\cD'(E)$) is square $\m$-integrable for all $q\in Q$.
Moreover, the correlation functions can be computed by the formula
\be
\lng q_1\cdots q_n\rng=i^{-n} \frac{\dr}{\dr\al^1}
\cdots\frac{\dr}{\dr\al^n}Z(\al^iq_i)|_{\al^i=0}= \int\lng
q_1,w\rng\cdots\lng q_n, w \rng \mu(w).
\ee
In particular, an integral of the function $\wt\f$ (\ref{f30})
over $\cD'(E)$ reads
\be
\int \wt\f\mu(w)=\int\lng\f'w\rng\m(w)=i\frac{\dr}{\dr
\al}Z(\al\f).
\ee

\bigskip
\bigskip
\noindent {\bf Appendix}
\bigskip

Let $\cK(E,\Bbb C)$ be the space of continuous complex functions
of compact support on $E=\Bbb R^n$. For each compact subset
$K\subset E$, we have a seminorm
\be
p_K(\f)=\op\sup_{x\in K} |\f(x)|
\ee
on $\cK(E, \Bbb C)$. These seminorms provide $\cK(E,\Bbb C)$ with
the topology of compact convergence. At the same time, $\cK(E,\Bbb
C)$ is a Banach space with respect to the norm
\be
\|f\|=\op\sup_{x\in E} |\f(x)|.
\ee
Its normed topology, called the topology of uniform convergence,
is finer than the topology of compact convergence. The space
$\cK(E,\Bbb C)$ can also be equipped with another topology, which
is especially relevant to integration theory. For each compact
subset $K\subset E$, let $\cK_K(E,\Bbb C)$ be the vector subspace
of $\cK(E, \Bbb C)$ consisting of functions of support in $K$. Let
$\cU$ be the set of all absolutely convex absorbent subsets $U$ of
$\cK(E,\Bbb C)$ such that, for every compact $K$, the set $U\cap
\cK_K(E,\Bbb C)$ is a neighborhood of the origin in $\cK_K(E,\Bbb
C)$ under the topology of uniform convergence on $K$. Then $\cU$
is a base of neighborhoods for the inductive limit topology on
$\cK(E,\Bbb C)$ \cite{rob}. This is the finest topology such that
the injection $\cK_K(E,\Bbb C)\to \cK(E,\Bbb C)$ is continuous for
each $K$. The inductive limit topology is finer than the topology
of uniform convergence and, consequently, the topology of compact
converges. The space $M(E,\Bbb C)$ of complex measures on $E$ is
the topological dual of $\cK(E,\Bbb C)$, endowed with the
inductive limit topology. The space $M(E,\Bbb C)$ is provided with
the weak$^*$ topology, and $\cK(E,\Bbb C)$ is its topological
dual. The following holds \cite{bourb6}.

\begin{lem} \label{f5} \mar{f5} Let $\ve_x$ denote the Dirac
measure of support at a point $x\in E$.  The assignment
\mar{f6}\beq
s_\ve:E\ni x\to \ve_x\in M(E,\Bbb C) \label{f6}
\eeq
is a homeomorphism of $E$ onto the subset $T_\ve\subset M(E,\Bbb
C)$ of Dirac measures endowed with the relative topology.
\end{lem}

Of course, $\cD(E) \subset \cK(E,\Bbb C)$, but the standard
topology of $\cD(E)$ is finer than its relative topology as a
subset of $\cK(E,\Bbb C)$. Let $\cD_R(E)$ denote $\cD(E)\subset
\cK(E,\Bbb C)$ provided with the relative topology, and let
$\cD'_R(E)$ be its topological dual endowed with the weak$^*$
topology. Then $M(E,\Bbb C)$ is homeomorphic to a subspace of
$\cD'_R(E)$ provided with the relative topology. At the same time,
$\cD'_R(E)$ is a subspace of $\cD'(E)$ endowed with the relative
topology. Thus, we have the morphisms
\be
E\ar^{s_\ve} M(E,\Bbb C)\ar \cD'_R(E)\ar \cD'(E),
\ee
whose composition leads to the homeomorphism $x\to\ve_x=\dl_x
d^nx\to \dl_x$ (\ref{f2}).

\end{document}